\documentclass[%
reprint,
superscriptaddress,
 amsmath,amssymb,
 aps,
prb,
]{revtex4-2}

\usepackage{graphicx}
\usepackage{dcolumn}
\usepackage{bm}
\usepackage{natbib}

\begin{document}

\preprint{APS/123-QED}
\title{Unconventional temperature evolution of quantum oscillations  in Sn-doped  Bi$_{1.1}$Sb$_{0.9}$Te$_{2}$S topological insulator }%
\author{Bruno Gudac}
\affiliation{Department of Physics, Faculty of Science, University of Zagreb, 10000 Zagreb, Croatia}
\author{Petar Sačer}
\affiliation{Department of Physics, Faculty of Science, University of Zagreb, 10000 Zagreb, Croatia}
\author{Filip Orbanić}
\affiliation{Department of Physics, Faculty of Science, University of Zagreb, 10000 Zagreb, Croatia}
\author{Ivan Kokanović}
\affiliation{Department of Physics, Faculty of Science, University of Zagreb, 10000 Zagreb, Croatia}
\author{Zoran Rukelj}
\affiliation{Department of Physics, Faculty of Science, University of Zagreb, 10000 Zagreb, Croatia}
\author{Petar Popčević}
\affiliation{Institute of Physics, 10000 Zagreb, Croatia}
\author{Luka Akšamović}
\affiliation{Institute of Solid State Physics, TU Wien, 1040 Vienna, Austria}
\author{Neven Ž. Barišić}
\affiliation{Department of Physics, Faculty of Science, University of Zagreb, 10000 Zagreb, Croatia}
\affiliation{Institute of Solid State Physics, TU Wien, 1040 Vienna, Austria}
\author{Munisa Nurmamat}
\affiliation{College of Physics and Electronic Information Engineering, Zhejiang Normal University, Jinhua, Zhejiang 321004, China}
\affiliation{Zhejiang Institute of Photoelectronics, Jinhua, Zhejiang 321004, China}
\author{Akio Kimura}
\affiliation{Graduate School of Advanced Science and Engineering,
Hiroshima University
1-3-1 Kagamiyama, Higashi-Hiroshima 739-8526,
Japan}
\affiliation{International Institute for Sustainability with Knotted Chiral Meta Matter (WPI-SKCM$^2$), Hiroshima University, Higashi-Hiroshima 739-8526, Japan}
\author{Mario Novak}
\email{mnovak@phy.hr}
\affiliation{Department of Physics, Faculty of Science, University of Zagreb, 10000 Zagreb, Croatia}

\date{\today}

\begin{abstract}
Among various topological insulators, Sn-doped Bi$_{1.1}$Sb$_{0.9}$Te$_{2}$S stands out for its exceptional properties. 
It has a wide energy gap and typically exhibits a well-isolated Dirac point and a Fermi level positioned within the gap. 
The samples we present display metallic-like low-temperature resistivity attributed to surface states, pronounced quantum oscillations observable even at 40 K, and a Fermi level located approximately 100 meV above the Dirac point.
In this work, we report an unusual effect: a strong temperature dependence of the quantum oscillation frequency, which decreases by around 10\% between 2 and 40 K. 
This reduction significantly exceeds the expected effects of the Sommerfeld and topological corrections for Dirac quasi-particles, which could account for only one-eighth of the observed change. We attribute this change to the temperature-induced renormalization of the bulk band gap size due to electron-phonon interactions, which in turn affect the position of the surface Dirac point within the gap. Furthermore, we propose that in this compound, surface quantum oscillations can serve as a precise tool for investigating the low-temperature evolution of the bulk band gap size.
\end{abstract}

\maketitle


\section{Introduction}
The quest to understand and harness exotic electronic phenomena in condensed matter systems has recently led to significant advancements in the field of topological insulators (TIs). 
The TIs are characterized by robust metallic surface states (SSs) located in the bulk band gap, and thus of great interest for (potential) applications in spintronics, quantum computing, and low-power electronics \cite{Qi2011, Fu2008, He2022}.

Utilizing numerical methods and topological quantum chemistry formalism, researchers have identified numerous compounds as topological insulators \cite{Vergniory2019}. 
However, only a few of those compounds have emerged as practical candidates for exploiting their unique topological properties. 
Among them, the second generation of Bi-based tetradymite compounds, described by the general formula (Bi$_{1-x}$Sb$_x$)$_2$(Se$_{1-y}$Te$_y$)$_3$, shows promising potential for leveraging quantum phenomena associated with topological surface states. 
These compounds demonstrate insulating behavior at high temperatures and crossover to a metallic-like state at lower temperatures, with the Dirac point and Fermi level residing within a wide bulk band gap \cite{Ren2010, Taskin2011, Arakane2012}.
\begin{figure*}[t]
\includegraphics[width=15cm]{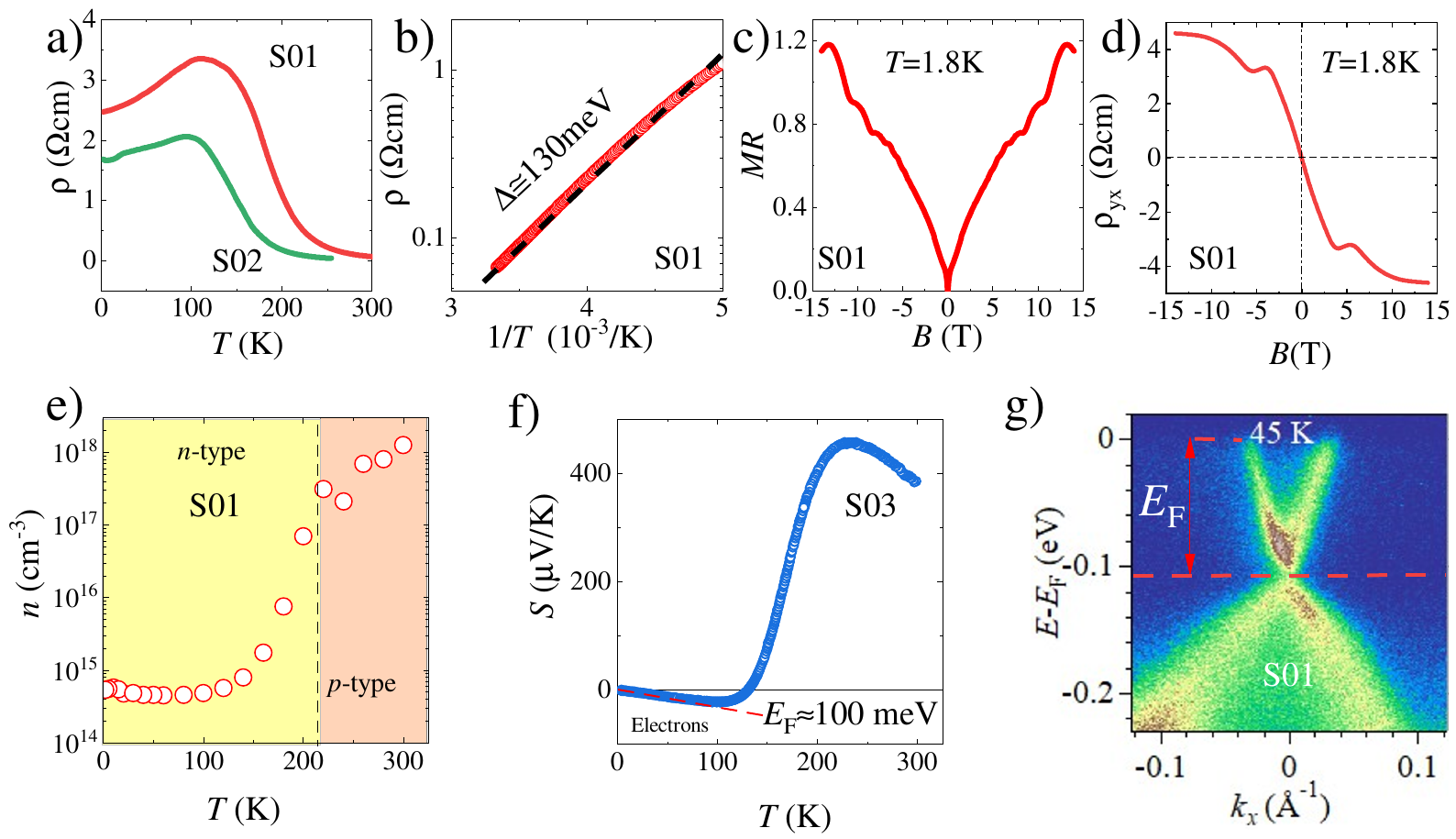}
\caption{\label{fig1} General sample characterization:
a) The temperature dependence exhibits the characteristic signature of a topological insulator with the Fermi level ($E_F$) located in a band gap.  At low temperatures, the behavior is metallic-like, attributed to dominant surface conductivity. b) At higher temperatures, the dependence resembles activation behavior with an activation energy of $\Delta \approx 130$ meV for sample S01.
c) The magnetoresistance ($MR$)  for sample S01 at 1.8 K shows notable quantum oscillations superposed on a weak-antilocalization-like signal. 
d) $\rho_{yx}$ as a function of $ B$ at 1.8 K for sample S01. In the  high-$B$ regime, the Hall signal shows non-linear $n$-type  behavior. 
e) The effective charge concentration is extrapolated from the low-field Hall signal. The charge concentration exhibits a swing from $p$-type to $n$-type as the temperature decreases, with a value of $5.1\times 10^{14}$ cm$^{-3}$ at 1.8 K, indicating good low-temperature charge compensation. 
f) The Seebeck effect for sample S03 displays a sharp drop in strength with a sign change near the $p$-type to $n$-type transition observed in sample S01. The temperature dependence is linear at low temperatures with a slope of $S/T = 0.26$ $\mu$V/K. Using  Mott formula, we estimate the position of $E_{F} \approx 100$ meV.
g) ARPES spectra of the Dirac surface states for sample S01. The Fermi energy is located around 100 meV above the Dirac point.   }
\end{figure*}

The key prerequisite for exploiting the properties of topological insulators is the sufficiently high mobility of the SS. 
Detection of quantum oscillations (QOs)  is a commonly used method to confirm the high mobility of the SSs.
Surface-originating QOs have been detected in the early years of TI research but were very weak and could be, in most cases, only detected after applying the derivative to the signal \cite{Ren2010, Xiong2012}. 
This points to the relatively low mobility of the SS. Here we investigate Sn-doped topological insulator Bi$_{1.1}$Sb$_{0.9}$Te$_2$S (BST2S),  a compound that exhibits well-pronounced surface QOs and one of the biggest gap among the tetradymite family, reaching 350 meV \cite{Kushwaha2016}.

This compound shows superior properties in comparison to prototypical TIs like Bi$_2$Se$_3$,  Bi$_2$Te$_3$, or some of the quaternary solid solutions (BiSb)$_2$(TeSe)$_3$, with the Dirac point being positioned in the gap and well-separated from the conduction and valence band thus allowing to study the SSs close to the Dirac point without the interface of the bulk states. 
The Dirac point separation and the increase in the energy gap are achieved by the higher electronegativity of the S atoms, reducing the absolute energy of the valence band and further reducing the lattice size by the introduction of Sb atoms \cite{Kushwaha2016, Yang2012, Teramoto1961}. 
Reports indicate that BST2S single crystals exhibit low-temperature decoupling of surface and bulk states, high-quality surface quantum oscillations,    and quantum Hall effect observed even on $\mu$m tick crystals \cite{Zhao2019, Xie2019, Ichimura2019a}. Reported robust surface state effects make BST2S an interesting candidate for studying various proximity-induced effects between TI and superconductors or ferromagnets. 

In this work, we investigate strong quantum oscillations in BST2S that are directly observable even at temperatures around 40 K. Interestingly, these surface QOs exhibit an unconventionally strong quadratic dependence of frequency on temperature. 
The observed overall change in frequency with temperature surpasses the theoretically predicted contributions for metallic bands, which arise from the temperature dependence of the chemical potential (the Sommerfeld term) and the linearity of the band dispersion (the topological term), by nearly an order of magnitude. 
In addition to the quantum oscillation analysis, we provide a detailed transport characterization that indicates the dominance of surface channels in low-temperature charge transport effects and confirms the high quality of the samples.

\section{Experimental}
Single crystals of doped Bi$_{1.1}$Sb$_{0.9}$Te$_2$S (where around 2 $\%$ of Bi is replaced with Sn) were grown using a modified Bridgman method. 
The stoichiometric amount of high-purity elements was mixed in a glove box, and the mixture was then sealed under an argon pressure of 1 bar.
The mixture was homogenized at 950 ° C for 48 hours, with thorough mixing performed at regular intervals.
Crystal growth was achieved by slowly cooling down to 500°C and annealing at this temperature for several days. 
The obtained material was characterized using XRD and energy-dispersive XRD to confirm its structure and composition.
Electrical transport measurements were conducted using a low-temperature cryostat equipped with a superconducting magnet and a variable temperature insert on samples with a 6-contact configuration made using silver paint.\footnote{Samples S01 and S02 were measured using two different superconducting magnet systems.} 
The thickness of the samples for transport measurements (S01 and S02) was approximately 20 $\mu m$. The Seebeck coefficient was measured on 100 $\mu m$ thick sample S03 from the same batch using Constantan-Au(0.08\% Fe) thermocouples to measure the temperature gradient and voltage difference. 
The measurements were performed using the "seesaw heating method" with heating elements made from thin-film resistors.
Angle-resolved photoemission spectroscopy (ARPES) was conducted at an energy of 46 eV on sample S01 (the same sample used for transport measurements).

\section{Result and discussion}
Detailed characterization of  BST2S  is given in Fig.\ref{fig1}. Temperature dependence of electrical resistivity (Fig.\ref{fig1}a) of two samples (S01 and S02) from the same batch exhibits high-temperature insulating-like (activation) behavior followed by low-temperature metallic-like behavior. 
The maximum in resistivity is observed at the crossover between high and low-temperature regimes ranging from 100 K and 120 K and is commonly associated with the onset of surface dominant transport \cite{Cai2018, An2018}. 
The Arrhenius plot for S01, in Fig.\ref{fig1}b, indicates the Fermi level to be located in the gap around $\Delta=130$ meV below the conduction band \cite{Kushwaha2016}. 
The magnetoresistance $MR=(\rho_B-\rho_0)/\rho_0$ at the lowest measured temperature  (Fig.\ref{fig1}c) shows strong field dependence with a pronounced signature of the Shubnikov-de Haas (SdH) quantum oscillations.
The Hall effect at low temperatures and low magnetic fields exhibits a negative slope, indicating dominance by $n$-type carriers (Fig.\ref{fig1}d). In higher fields, we see the nonlinear behavior coming from the presence of the multi-channel effects - surface and bulk contributions. In addition, the low-$B$ Hall measurements as a function of temperature for sample S01 in Fig.\ref{fig1}e show the transition from $p$-type to $n$-type-dominated transport by lowering the temperature,   a characteristic commonly observed in the insulating members of the tetradymite family \cite{Kushwaha2016,Ren2010, Taskin2011}. 
Estimated  low-$T$ effective 3D charge concentration using the single band model is of the order of $5 \times 10^{14}$ cm$^{-3}$ showing good bulk charge compensation.
The Seebeck coefficient of sample S03  (Fig.\ref{fig1}f) is positive at high-$T$ and reaches relatively high values with the maximum value of   450 $\mu$V at 230 K. 
Below 100 K the Seebeck coefficient is negative and linearly approaches zero,  a hallmark of metallicity with electrons as dominant carriers. By applying the single band Mott formula:
\begin{equation}
    S_{Mott}=(-\pi^2/3e)k_b^2T/E_{F},
\end{equation}
which is still applicable for the Dirac systems in the limit of $E_F>>k_bT$ \cite{Ivanov2018a},  gives the value of the Fermi energy of $100$ meV.  
Finally, the ARPES measurements (Fig.\ref{fig1}g) show Dirac-like surface states with the Fermi energy located close to 100 meV above the Dirac point, implying that the low-$T$ Seebeck effect is indeed dominated by the metallic surface states. 

Keeping this in mind, we can analyze the Hall effect in the limit of low-$ T$ and low-$B$ in the light of the surface-dominant transport. Multiplying the bulk effective charge concentration with the sample thickness of 20 $\mu $m for the S01 sample, we get the effective surface charge concentration of $1.0(1)\times 10^{12}$ cm$^{-2}$. 
This value can be compared with the theoretically expected  Dirac surface charge concentration:
\begin{equation}
   n_s=\frac{1}{2\pi}\int_{0}^{E_F} \frac{E}{(v_f \hbar)^2}\,dE. 
\end{equation}
Using the ARPES obtained parameter for the Fermi energy $E_F=100$ meV and Fermi velocity $v_F=4.4 \times 10^{5}$ m/s (slope of the linear Dirac dispersion) we get  $n_s=1.1\times 10^{12}$ cm$^{-2}$, which closely matches the values obtained from the Hall effect.
To understand this matching, we have to look at the two-band Hall model in the low-$B$ limit. In the low-$B$ limit, the Hall signal is linear in $B$, and it depends on the charge concentration and mobility of each contributing channel. If the surface channel mobility strongly surpasses the bulk channel, the bulk channel would have a negligible contribution to the total Hall signal in the  low-$T$, low-$B$ limit. 

\begin{figure}
\includegraphics[width=8.5cm]{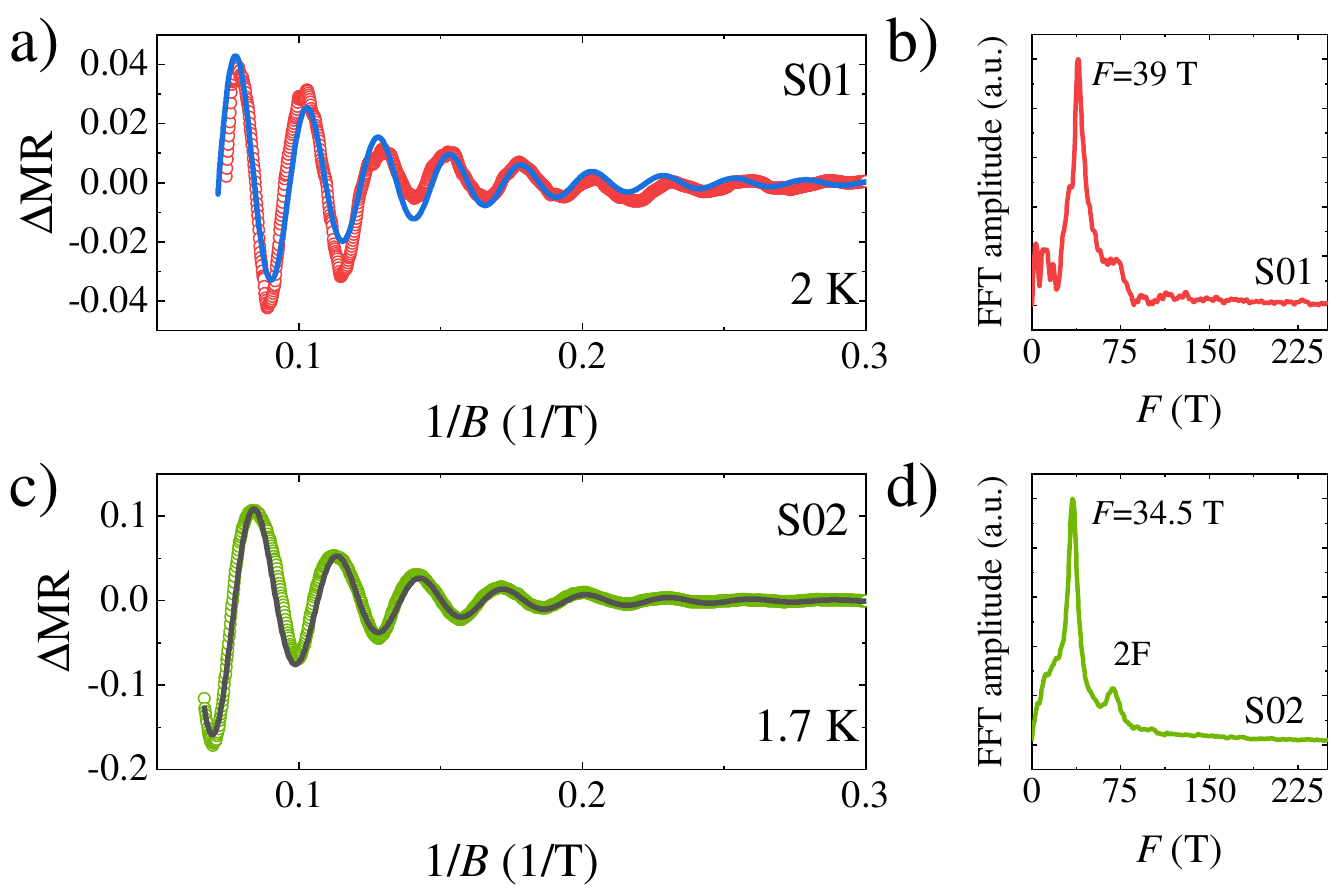}
\caption{\label{fig2}  a) Extracted SdH quantum oscillations for sample S01 at 2 K (red open symbols) with the L-K fit to the data using the mean-square method (blue line). Associated fast Fourier transform spectra showing single-frequency oscillation is given in panel b).
c) SdH quantum oscillations for sample S02 at 1.7 K (green open symbols) with the L-K fit (black line) and its fast Fourier transform spectra showing single-frequency oscillations with visible first and second harmonics in panel d. }
\end{figure}

\begin{figure*}
\includegraphics[width=18cm]{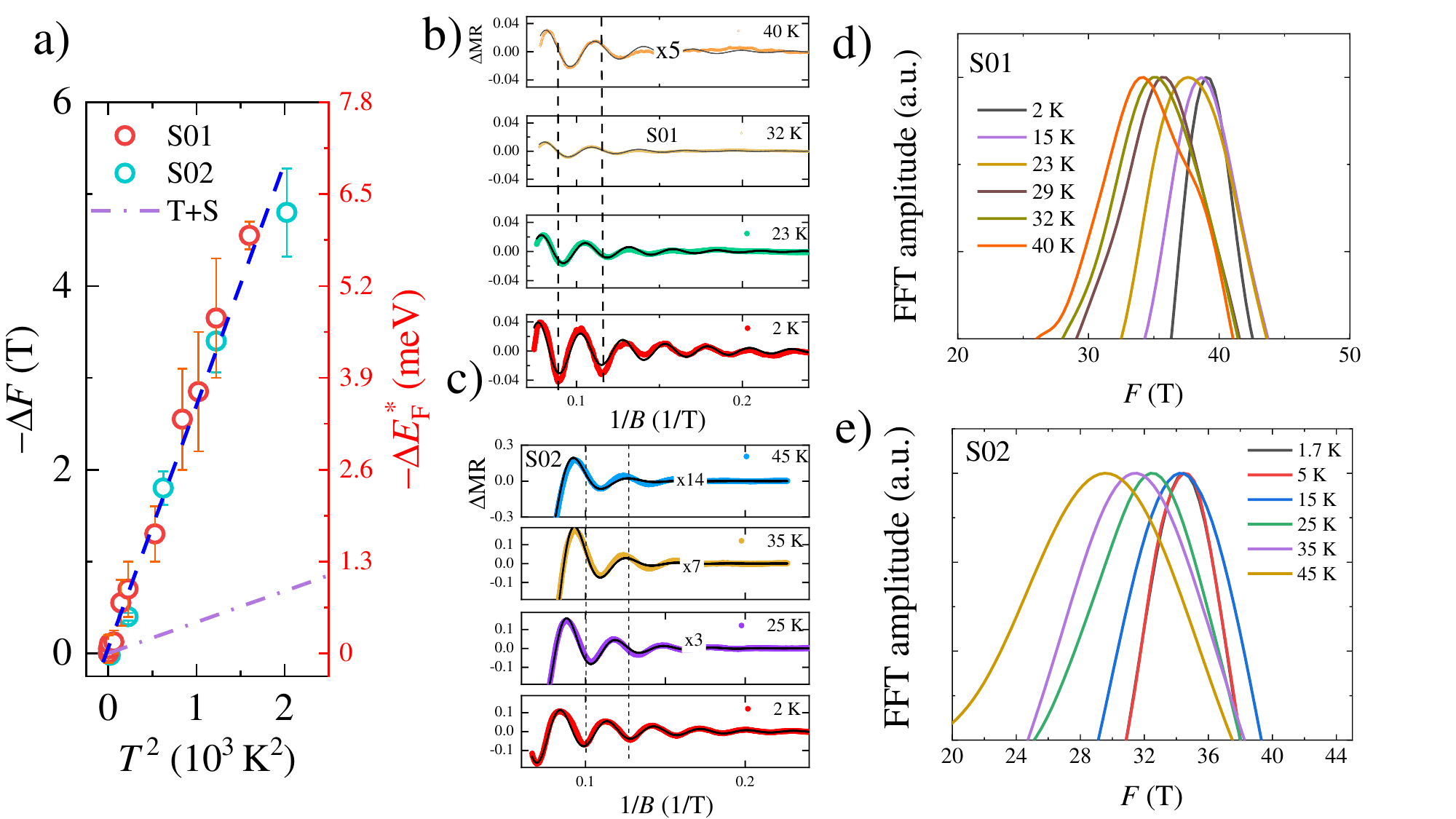}
\caption{\label{fig3} a) Temperature-dependent quantum oscillation frequency shift. The shift shows quadratic dependence with the slope of 0.0027(2) T/K$^2$ (the dashed line). The dash-dot line represents the expected combined   Sommerfeld and topological contribution to the temperature dependence of the SdH oscillation frequency for the Dirac dispersion ($m_{c}=0.1 m_{e}$ and  $F=39$ T). The right scale $\Delta E^{*}_F$ presents a corresponding frequency shift in units of energy for sample S01 by using Eq.\ref{eq3}. b), c) Field-dependent SdH quantum oscillations for samples S01 and S02 clearly show a change in the frequency by increasing temperature. d) and e) FFT analysis of the oscillations for S01 and S02. In both samples, the  FFT peaks move to lower values by increasing temperature.}
\end{figure*}

Next, we turn to the analysis of the QOs in magnetoresitance.  The SdH quantum oscillations of samples S01 and S02 given in Fig.\ref{fig2} were extracted by subtracting a smooth background from the symmetrized data. 
Both samples show single-frequency oscillations with frequencies 39.0(3) T for S01 and 34.5(3) for S02 at around 2 K. 
The higher quality of the extracted oscillations in the case of S02 allows for the detection of a second harmonic.
In principle, the frequency of the quantum oscillations can be connected to charge concentration. 
If we assume that the Fermi surface responsible for QOs is 2D and isotropic, then the charge carrier density is given by $n_s=k_F^2/4\pi$ for the case of spin-polarized Dirac surface states. 
Furthermore, according to the Onsager relation, $k_F$ (a Fermi wave vector) and Fermi energy $E_F$ are related to the frequency of quantum oscillation:
\begin{equation}\label{eq3}
    F=\hbar/(2\pi e)\pi k_F^2=1/(2e\hbar)(E_F/v_F)^2. 
\end{equation}
For the sample S01 surface charge concentration obtained from the quantum oscillations is  $n^{SdH}_s=0.97 \times 10^{12}$ cm$^{-2}$   and  the Fermi energy is $E_F=\hbar v_F k_F= 100$ meV. These values are in excellent agreement with independently obtained values from the Seebeck effect, ARPES, and low-$B$ Hall effect, affirming our assumption that the QOs are coming from the surface and that the surface dominates the charge transport effects at low temperatures.  

As a further step of the analysis, we fit the Lifshitz-Kosevich (L-K) formula to experimental data (Fig. \ref{fig2}a  and \ref{fig2}c). We use a modified expression
\footnote{For the L-K fit, we utilized the standard expression for a 2D system \cite{shoenberg_1984}. We reduced the number of fitting parameters, as described in the main text, based on the justified assumption that the oscillations originate from the Dirac surface states in order to provide more readable information on the frequency. The expression used is:
\begin{equation}
MR=C\sqrt{B/F}A_DA_T\cos(2\pi (F/B -1/2)+ \phi_B)
\end{equation} 
where $C$ is a amplitude coefficient, $A_T=2\pi^2(k_B T/v_F B)\sqrt{2F/e\hbar}/\sinh(2\pi^2(k_B T/v_F B)\sqrt{2F/e\hbar})$, $A_D=2\pi^2(k_B T_D/v_F B)\sqrt{2F/e\hbar}$, the Berry phase $\phi_B$ is fixed to $\pi$ which makes the total phase factor to vanish. $k_B$ is Boltzmann constant, $\hbar$ reduced Planck constant, $e$ electron charge, $v_F$ is Fermi velocity set from the ARPES measurements to $4.4x10^{5}$ m$/$s, $T_D$ is the Dingle temperature.}
where the cyclotron mass is not a free-fitting parameter but is related to the Fermi energy (i.e.,  frequency of the oscillations) and the Fermi velocity \footnote{It is assumed that the Fermi velocity is the same for both samples and corresponds to the ARPES obtained value for the sample S01.} $m_c=E_F/v^2_F=\sqrt{2e\hbar F}/v_F$.\cite{Guo2021, shoenberg_1984}   
This way, we have maximally reduced the number of fitting parameters. Frequency $F$ was kept as a fitting parameter, although, in principle, it could be fixed from the peak position of the fast Fourier transform (FFT) spectra. 
The frequency was not fixed in order to have an additional independent estimation of its value, which will be relevant in further discussion for oscillations at higher temperatures. The L-K fit frequencies are very close in values to the FFT frequencies for both samples.  
Besides the oscillation frequency, the L-K fit gives us information on the quantum scattering time $\tau_q$, which can be related to the mobility of the surface states by using the expression $\mu_{SdH}=e\tau_q/m_c$. For sample S01, the surface quantum mobility is around 2200 cm$^2$/Vs, and for S02  2000 cm$^2$/Vs \footnote{Taking into account the size of the activation energy for S01 of 130 meV we can practically neglect the contribution of the bulk channel at 2K and make an estimation of the surface Drude mobility to be $\mu_s=\rho/(t e n_{s})=5100$ cm$^2$/Vs, where $t$ is the sample thickness.}. These values are comparable to previously reported values and smaller than seen in vanadium-doped BST2S, which shows QOs up around  50 K\cite{Zhao2019}. Finally, using the oscillation frequencies, the effective mass for S01 is estimated to be  $m_c=0.1 m_e$ and $0.095 m_e$ for S02.

The surface charge carriers' low effective mass and high mobility result in pronounced quantum oscillations that can be detected even at around 40 K for both presented samples. 
A closer inspection of the oscillations' temperature evolution revealed that the oscillation frequency has a temperature dependence where the increase in temperature lowers the frequency. Figure \ref{fig3}a  shows almost identical quadratic $T$ dependence for samples S01 and S02, with the average value of the quadratic term of 0.0027(2) T/K$^2$. The quadratic $T$ dependence can be regarded as a first-term in a perturbation correction expansion. 
The temperature dependence of quantum oscillation frequency is not a commonly observed effect  \cite{Nowakowska2023}. 
Thus, we checked the frequency $T$ dependency by two independent procedures: the FFT and L-K fitting. Both methods give the same results, and the values in Fig.\ref{fig3}a are presented as the mean value of these two methods for a particular temperature. The observed shift is not a consequence of background subtraction, and the shift in the position of the oscillation peaks can also be observed even in the raw magnetoresistance data \footnote{For the background subtraction, we used a smooth polynomial curve of the 3rd order}. To support our claims in Fig.\ref{fig3}b and \ref{fig3}c, we show an oscillating part of magnetoresistance for samples S01 and S02 at different temperatures. From the figures, clearly, a change in the pattern of the oscillations can be observed with increasing temperature.   The FFT spectra given in Fig.\ref{fig3}d and  \ref{fig3}e    show that the change in the oscillation pattern originates from the temperature-induced shift of the oscillation frequency.

As a first step in the analysis, we examine the contribution of the Sommerfeld correction arising from the temperature dependence of the chemical potential for the 2D Dirac system. The quadratic term is equal to
\begin{equation}
    \Delta F_S=-\frac{1}{12e \hbar}\left(\frac{\pi k_B}{v_f}\right)^2 T^2,
\end{equation} 
giving correction factor of the order $5\times10^{-5}$ T/K$^2$. 
Recently, it was realized that the Dirac-like system can have an additional topological contribution due to an energy dependence of the cyclotron mass \cite{Guo2021}. 
This additional contribution has been successful in describing the temperature dependence of quantum oscillations in Cd$_3$As$_2$ and CaFeAsF \cite{Guo2021, Terashima2022}.
The topological term is equal to:
\begin{equation}
    \Delta F_T=-\frac{1}{2e \hbar}\left(\frac{\pi k_B}{v_f}\right)^2 T^2
\end{equation}
giving correction factor of the order $3\times10^{-4}$ T/K$^2$. 
If we combine Sommerfeld and topological corrections, our observed effect is still approximately eight times stronger.  

Given that the topological contribution does not account for the observed frequency shift, we need to explore other possible explanations. One possibility is that the QO frequency shift is attributed to presumed surface band bending. In some instances of TIs, it has been observed that surface band bending evolves over time, altering the surface states from $p$-type to $n$-type and effectively lowering the Dirac point in energy \cite{Taskin2011}. However, this effect does not explain our experimental findings. Firstly, in our case, the Dirac point moves up in energy, and secondly, the frequency shift we observe is reversible with temperature.

The next candidate is the renormalization of the bulk band gap.
It is well-established that the renormalization of the bulk band gap can occur due to electron-phonon interaction effects, and this process is temperature-dependent. The simplest correction in this picture is proportional to the thermal volume expansion coefficient of the crystal lattice \footnote{From the powder X-ray diffraction  data, we got lattice parameters at   12 K $a=0.41910(8)$ nm $c=2.9476(6)$ nm and at 300 K $a=0.42029(7)$nm $c=2.9566(5)$nm which gives a thermal expansion coefficient of  $\Delta V/V= 0.3 \%$.}
and higher corrections are related to the self-energy diagrams of the electron-phonon interaction  \cite{Bardeen1950, Cardona2005}. 
For the system with an inverted band gap, it is common for the band gap to red-shift (decrease in size) with increasing $T$. 
This change in the energy position of the valence and conduction bands will consequently affect the position of the Dirac point of the surface states.
Under a reasonable assumption, the Fermi level is pinned by the Sn resonant impurities located in the band gap, a common occurrence in semiconductors \cite{Skipetrov2014}. We can indirectly get information on the band gap temperature evolution since the frequency shift, $\Delta F$, directly provides information about the distance between the Fermi level and the Dirac point, whose position depends on the energy of the valence and conduction band.

In Fig.\ref{fig3}a, the frequency shift for sample S01 is recalculated in the units of energy. 
The energy shift between 2 K and 45 K is approximately 6.5 meV. 
This can be compared with the previously reported temperature shift of the Van Hove singularity (VHS) peak observed in optical spectroscopy, which is attributed to optical transitions at the center of the Brillouin zone. 
The VHS peak, with an increasing temperature from 5 K to 50 K, exhibits a red-shift of around 7.0(5) meV  \cite{Jiang2020}, which closely aligns with our quantum oscillation energy shift.

It's noteworthy that the rigorous theoretical modeling predicts  $T^4$ dependence of the gap size in the asymptotic limit of low temperatures and linear behavior at high temperatures \cite{Cardona2004, Cardona2005}. 
However, our observed temperature dependence clearly does not reflect this prediction.
We could speculate that we are far from the asymptotic low-temperature limit for the model to be applicable. 
This speculation is supported by the often observed quadratic gap temperature behavior in the intermediate temperature range of semiconductors \cite{Varshni1967}.

\section{Conclusion}
We have studied Sn-doped Bi$_{1.1}$Sb$_{0.9}$Te$_{2}$S topological insulator. By detailed characterization, we confirmed the excellent sample quality characterized by low effective bulk charge concentration and strong surface quantum oscillations.
Moreover, the surface dominates the low-temperature resistivity, Seebeck, and low-field Hall effect. 
By using the Seebeck effect, ARPES, and quantum oscillation results, the Fermi energy is positioned in the gap at 100 meV above the Dirac point. 
Interestingly, we have observed an unconventionally strong temperature dependence of the quantum oscillation frequency.     
The frequency temperature dependence is quadratic with the strength of 0.0027(2) T/K$^2$, and it is observable up to 40 K.
It cannot be satisfactorily explained by using the temperature dependence of the chemical potential or topological correction due to the energy dependence of the effective mass.  A comparison of the energy scale of the temperature-induced frequency shift and previously reported optical measurements of the Van Hove singularity indicates that the observed temperature dependence is closely related to the dynamics of the energy of the bulk band gap. 
Hence, the obtained QO temperature dependence could be utilized as a valuable and precise indirect tool to gain insight into the evolution of the band gap of BST2S at low temperatures.


\section*{Acknowledgements }
\textbf{Acknowledgements:} This work was supported by the Croatian Science Foundation  projects  IP 2018 01 8912,  IP-2022-10-3382  and IP-2020-02-9666, the CeNIKS project co-financed by the Croatian Government and the EU through the European Regional Development Fund–
Competitiveness and Cohesion Operational Program (Grant
No. KK.01.1.1.02.0013) and FWF Project P 35945-N. APRES experiments were conducted at BL01 of HiSOR under Proposal No. 21G025

\bibliography{BST2S-Gudac}

\end{document}